\documentclass[12pt]{iopart}
\usepackage{graphicx}
\usepackage[utf8]{inputenc}
\usepackage[english]{babel}

\usepackage{iopams}  
\begin{document}

\def\aligned{\vcenter\bgroup\let\\\cr
\halign\bgroup&\hfil${}##{}$&${}##{}$\hfil\cr}
\def\endaligned{\crcr\egroup\egroup}

\title[Point Cloud Transformers applied to Collider Physics]{Point Cloud Transformers applied to Collider Physics}

\author{Vinicius Mikuni$^1$, Florencia Canelli$^1$}

\address{$^1$ University of Zurich,\\Winterthurerstrasse 190, 8057 Zurich, Switzerland}
\ead{vinicius.massami.mikuni@cern.ch}
\vspace{10pt}

\begin{abstract}
Methods for processing point cloud information have seen a great success in collider physics applications. One recent breakthrough in machine learning is the usage of Transformer networks to learn semantic relationships between sequences in language processing. In this work, we apply a modified Transformer network called Point Cloud Transformer as a method to incorporate the advantages of the Transformer architecture to an unordered set of particles resulting from collision events. To compare the performance with other strategies, we study jet-tagging applications for highly-boosted particles.
\end{abstract}
\maketitle
\section{Introduction}
\label{sec:intro}
The interactions between elementary particles is described by the Standard Model (SM) of particle physics. Particle colliders are then used to study these interactions by comparing experimental signatures to SM predictions. At every collision, hundreds of particles can be created, detected, and reconstructed by particle detectors. Extracting relevant physics quantities from this space is a challenging task that is often accomplished through the usage of physics-motivated summary statistics, that reduce the data dimensionality to manageable quantities. A recent approach is to interpret the set of reconstructed particles as points in a point cloud. Point clouds represent a set of unordered objects, described in a well-defined space, often used for applications in self-driving vehicles, robotics, and augmented reality, to name a few. With this approach, information from each bunch crossing in a particle collider is interpreted as a point cloud, where the goal is to use this high-dimensional set of reconstructed particles to extract relevant information. However, extracting information from point clouds can also be a challenging task. One novel approach is to use the Transformers architecture \cite{DBLP:journals/corr/VaswaniSPUJGKP17} to learn the semantic relationship between objects. Transformers yielded a great success in recent years applied to natural language processing (NLP), often showing a superior performance when compared to previous well-established methods. The advantage of this architecture is the capability of learning semantic affinities between objects without losing information during long sequences. Transformers are also easily parallelizable, a huge computational advantage over sequential architectures like gated recurrent \cite{DBLP:journals/corr/ChungGCB14} and long short-term memory \cite{DBLP:journals/corr/abs-1801-07829} neural networks.
Applications of the Transformer network have already been applied to examples outside NLP problems, with examples in image recognition \cite{wu2020visual,dosovitskiy2020image}.

However, the original Transformer architecture is not readily applicable to point clouds. In fact, since point clouds are intrinsically unordered, the Transformer structure has to be modified to define a self-attention operation that is invariant to permutations of the inputs. A recent approach introduced in \cite{guo2020pct} addresses these issues through the development of Point Cloud Transformer (PCT). In this approach, the input points of the point cloud are first passed through a feature extractor to create a high dimensional representation of the particles features. The transformed data is then used as an input to a self-attention module. This module introduces attention coefficients per particle that learn the relative pairwise importance of each point in the set.

In this work, we will first describe the key features of the PCT architecture, and then evaluate it in the context of a high energy physics task, in the form of jet-tagging. Due to the nature of our example task the segmentation module of the PCT is not required. Results are compared with other benchmark implementations using three different public datasets.

\section{Related works}
\label{sec:related}
Neural network architectures that treat collision events as point clouds have recently grown in number given their state-of-the-art performance when applied to different collider physics problems. A few examples of such applications are jet-tagging \cite{Komiske:2018cqr,Dolan:2020qkr}, secondary vertex finding \cite{Shlomi:2020ufi}, event  and object reconstruction \cite{Qasim:2019otl,Fenton:2020woz,Duarte:2020ngm,Ju:2020xty,Pata:2021oez}, and jet parton assignment \cite{Lee:2020qil}. A comprehensive review of the different methods is described in \cite{Shlomi:2020gdn}.

Two particular algorithms for jet-tagging will be relevant for the following discussions of the PCT implementation. These are the ParticleNet \cite{PhysRevD.101.056019} and ABCNet \cite{Mikuni:2020wpr} architectures. The former introduces the EdgeConv operation for jet tagging, initially developed in \cite{DBLP:journals/corr/abs-1801-07829}. This operation uses a k-nearest neighbors approach to create local patches inside a point cloud. The local information is then used to create high level features for each point that retains the information of the local neighborhood. ABCNet, on the other hand, uses the local information to define an attention mechanism that encoded the neighborhood importance for each particle. This idea was first introduced in \cite{velikovi2017graph} and applied in \cite{chen2019gapnet}. A similar concept of attention mechanisms are defined for PCT, where a self-attention layer is used to provide the relationship importance between all particles in the set.

Jet-tagging is a common task used to benchmark different algorithms applied to collider physics. While a number of algorithms have been proposed in recent years, special attention will be given to algorithms with results in public datasets. In \cite{PhysRevD.101.056019}, results are presented for both quark gluon and top tagging datasets, while \cite{Moreno:2019bmu} introduces a multiclassification sample containing five different jet categories. The description of each dataset is discussed in Sec.~\ref{sec:class}.

\section{PCT network}
\label{sec:net}
The Transformer implementation applied to point clouds requires two main building blocks: the feature extractor and the self-attention (SA) layers. The feature extractor is used to map the input point cloud $F_{in} \in \mathbb{R}^{N\times\mathrm{d_{in}}}$ to a higher dimensional representation F$_e$ $\in  \mathbb{R}^{N \times \mathrm{d_{out}}}$. This step is used to achieve a higher level of abstraction for each point present in the point cloud. In this work, two different strategies are compared. An architecture consisting of stacked, one-dimensional convolutional (Conv1D) layers, and a second option, based on EdgeConv blocks. The EdgeConv block consists of an EdgeConv operation \cite{DBLP:journals/corr/abs-1801-07829} followed by 2 two-dimensional convolutional (Conv2D) layers and an average pooling operation across all neighbors. To ensure the permutation invariance between particles is maintained, all convolutional layers are implemented with stride and kernel size of 1 and unless otherwise stated, are followed by a batch normalization operation and ReLU activation function. All convolutional layers are implemented such that the particle information is preserved, i.e for N particles with F features, the convolutional operations only modify F while N is unchanged. The EdgeConvolution operation uses a k-nearest neighbors approach to define a vicinity for each point in the point cloud. This enhances the ability of the network to extract information from a local neighborhood around each point.

The first strategy is referred to simple PCT (SPCT) while the second will be referred to just PCT.

The second main building block is the usage of an offset attention defined as a self-attention (SA) layer. The output of the feature extractor F$_e$ is used as the input of the first SA layer. The goal of the SA layer is to determine the relationship between all particles of the point cloud through an offset attention mechanism. This approach differs from the one taken in ABCNet, where a self-attention and neighboring attention coefficients are defined for a neighborhood of each particle.

In the same terms defined in the original Transformer \cite{DBLP:journals/corr/VaswaniSPUJGKP17} work, three different matrices, all built from linear transformations of the original inputs. These matrices are called query (Q), key (K), and value (V). Attention weights are created by matching Q and K through matrix multiplication. These attention weights, after normalization, represent the weighted importance between each pair of particles. The self-attention is then the result of the weighted elements of V, defined as the result of the matrix multiplication between the attention weights and the value matrix. 
The linear transformations are accomplished through the usage of Conv1D layers such that:
\begin{equation}
    \begin{aligned}
        \mathrm{Q, K, V} &= \mathrm{F}_e.\mathrm{(W_q, W_k, W_v)} \\
        \mathrm{Q, K} &\in \mathbb{R}^{N \times \mathrm{d_a}}, \mathrm{V} \in \mathbb{R}^{N \times \mathrm{d_{out}}}.
    \end{aligned}
\end{equation}
The dimension $\mathrm{d_a}$ of the Q and K matrices is independent from the dimension $\mathrm{d_{out}}$. In this work, and similarly to the choice used in the original PCT implementation, $\mathrm{d_a}$ is fixed to $\mathrm{d_{out}}/4$. While other options can be used, we have not observed further improvements for different choices of $\mathrm{d_a}$.
To maintain the linear operations in these definitions, no batch normalization or activation function are used.
The matrices $\mathrm{(W_q, W_k, W_v)}$ contain the trainable linear coefficients introduced by the convolutional operation. The attention weights (A) are then calculated by first multiplying the query matrix with the transpose of the key matrix followed by a softmax operation:

\begin{equation}
    \mathrm{A} = \mathrm{Softmax}(\mathrm{Q.K}^{\mathrm{T}})/N, \mathrm{A} \in \mathbb{R}^{N \times N}.
    \label{eq:A}
\end{equation}

The softmax operation is applied to each row of A to normalize the coefficients for all points, while the columns are normalized using the L1 norm (represented with N in the equation), after the Softmax operation. This decision follows the approach introduced in the PCT model, different from the original Transformer implementation that used $1/\sqrt{d}$, with $d$ the embedding dimension. 

The attention weights are then multiplied by the value matrix, resulting in the self-attention F$_{sa}$ with
\begin{equation}
    \mathrm{F}_{sa} = \mathrm{A\cdot V}, \mathrm{F}_{sa} \in \mathbb{R}^{N \times \mathrm{d_{out}}}.
    \label{eq:fsa}
\end{equation}

While the original Transformer architecture uses F$_{sa}$ to derive an absolute attention per object, it was observed by the PCT authors, and later revisited in Section \ref{sec:ablation}, that the usage of an \textit{offset-attention} might result in a superior classification performance.  The offset-attention defines the self-attention coefficients as per particle modifications of the inputs, rather than an absolute quantity per particle. To calculate the offset-attention, first the difference between F$_e$ and F$_{sa}$ is calculated as the initial offset. This initial offset is then passed through a Conv1D layer with same output dimension $\mathrm{d_{out}}$. The result of this layer is now the offset added to the initial inputs F$_e$.
Different levels of abstraction can be achieved by stacking multiple SA layers, using the output of each SA layer as the input for the next.

To complete the general architecture, the SA layers are combined through a simple concatenation over the feature dimension, followed by a mean aggregation, resulting in the overall means of each feature across
all the particles, differently from the original PCT implementation that instead uses a maximum pooling operation. The output of this operation is passed through fully connected layers before reaching the output layer, normalized through a softmax operation. Similarly to the convolutional layers, all fully connected layers besides the output layers are followed by a batch normalization and ReLU activation function.

\begin{figure}[htb]
    \centering
    \includegraphics[width=0.99\textwidth]{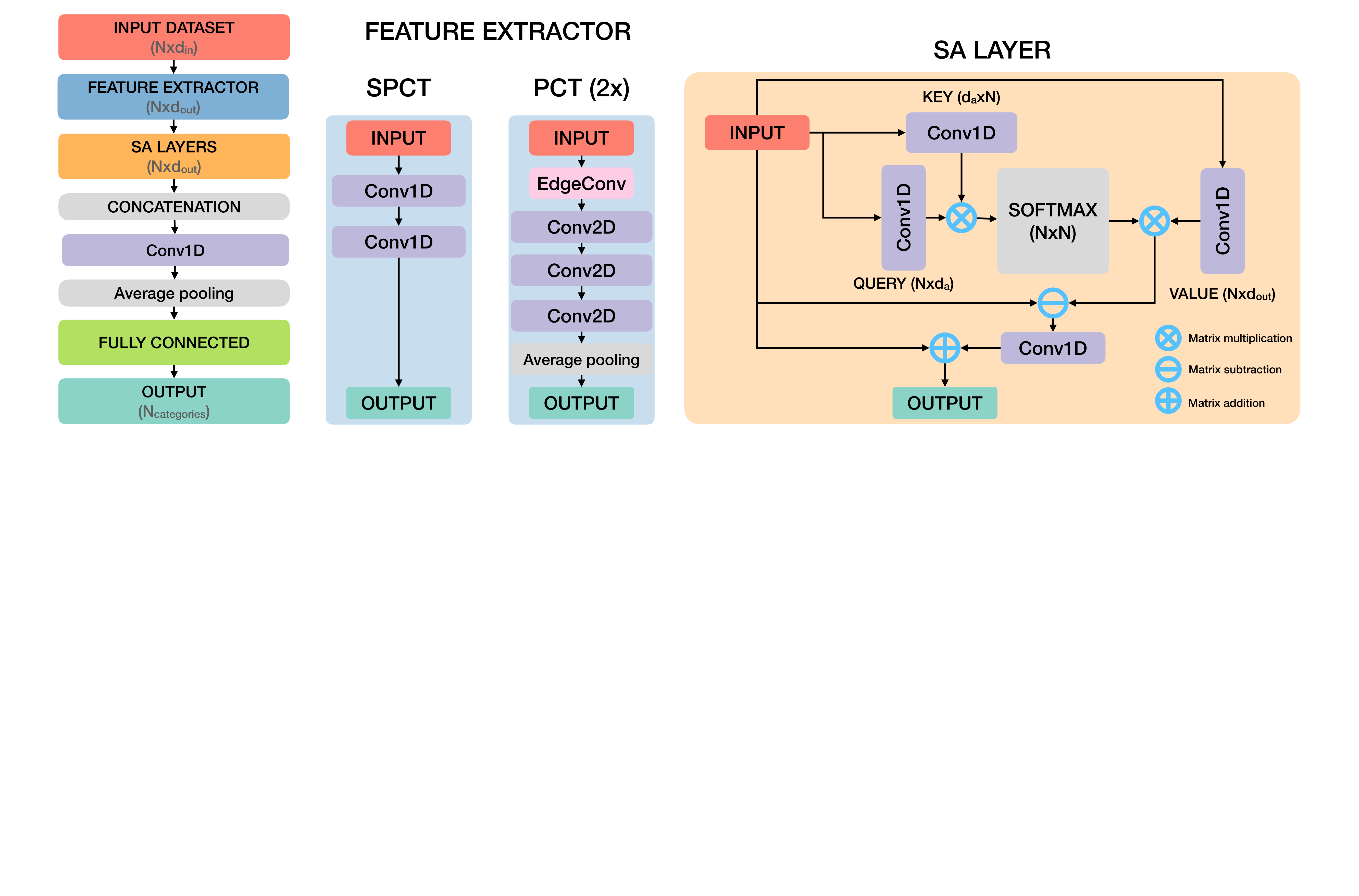}
    \caption{General network architecture (left), feature extractor (middle), and self-attention layer description (right). d$_{in}$, d$_{out}$, and d$_{a}$ represent the input and output feature sizes and the dimension used in the definition of the key and query matrices. The ReLu activation function, followed by a batch normalization operation are implied in all convolutional and fully connected layers, besides the ones used in the SA layers to define the linear transformations. }
    \label{fig:PCT}
\end{figure}

The general PCT network and the main building blocks are shown in Fig.~\ref{fig:PCT}. The training details are explained in Sec.~\ref{sec:training_details}

\section{Training details}
\label{sec:training_details}
The PCT implementation is done using Tensorflow v1.14 \cite{tensorflow2015-whitepaper}. A Nvidia GTX 1080 Ti graphics card is used for the training and evaluation steps. For all architectures, the Adam optimiser \cite{2014arXiv1412.6980K} is used with a learning rate starting from 0.001 and decreasing by a factor 2 every 20 epochs, to a minimum of 1e-6. The training is performed with a mini batch size of 64 and a maximum number of 200 epochs. If no improvement is observed in the evaluation set for 15 consecutive epochs, the training is stopped. The epoch with the lowest classification loss on the test set is stored for further evaluation.

\section{Jet classification}
\label{sec:class}
Different performance metrics are compared for (S)PCT applied to a jet classification task on different public datasets. Jets are collimated sprays of particles resulting from the hadronization and fragmentation of energetic partons. Jets can show distinguishing radiation patterns depending on the elementary particle that has initiated the jet. Traditional methods use this information to define physics-motivated observables \cite{Thaler:2010tr} that can distinguish different jet categories.

The PCT architecture uses two EdgeConv blocks, each defining the k-nearest neighbors of each point with k = 20. The initial distances are calculated in the  pseudorapidity-azimuth ($\eta-\phi$) space of the form $\Delta R = \sqrt{\Delta\eta^2 + \Delta\phi^2}$. The distances used for the second EdgeConv block are calculated using the full-feature space produced in the output of the last EdgeConv block.

\begin{figure}[htb]
    \centering
    \includegraphics[width=0.90\textwidth]{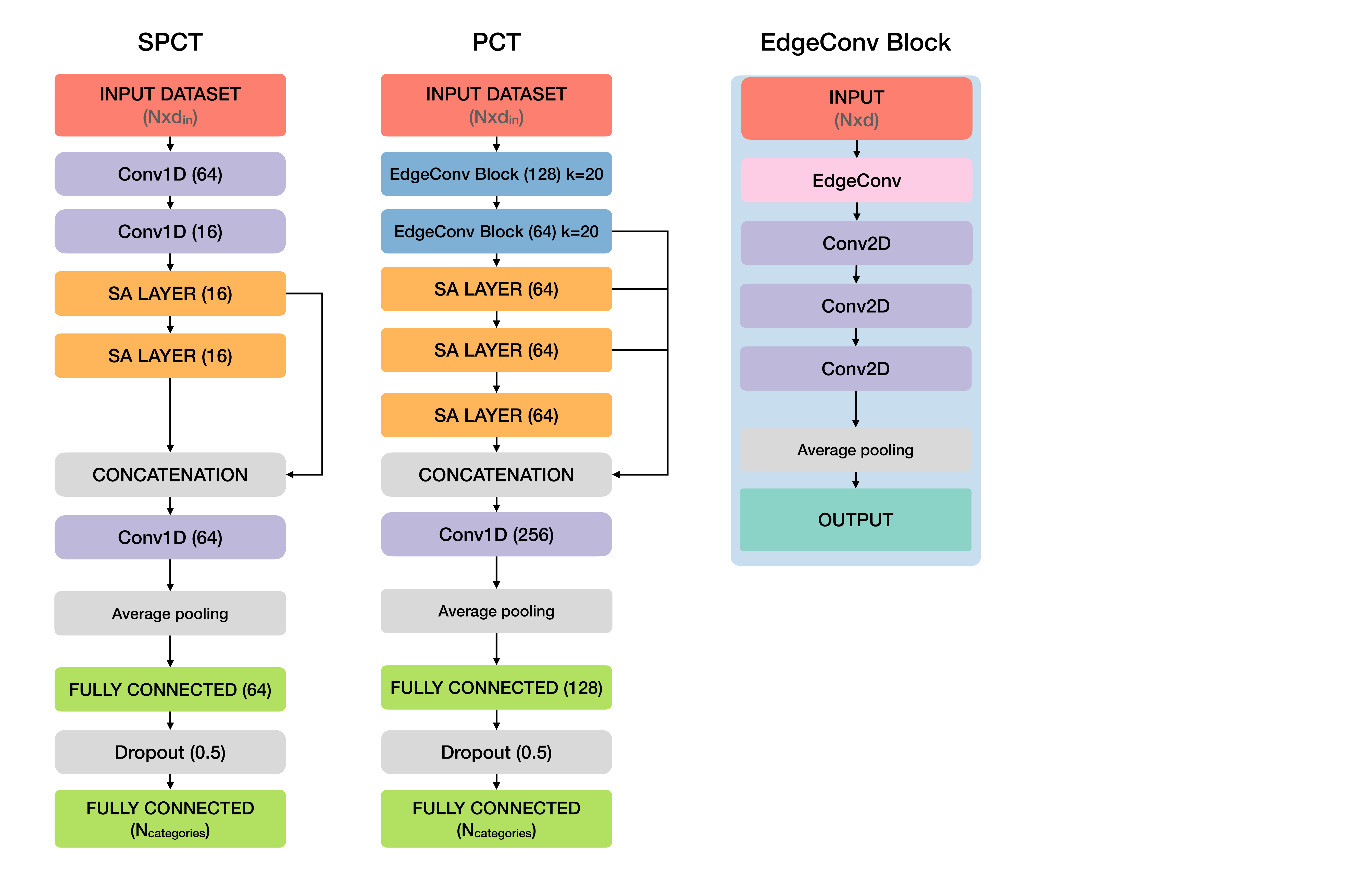}
    \caption{SPCT (left) and PCT (middle) architectures used for all jet-tagging classification tasks. The EdgeConv block structure used in PCT is shown in the right. Numbers in parenthesis represent layer sizes. }
    \label{fig:PCT_arc}
\end{figure}

Besides the feature extractor, PCT uses three SA layers while SPCT uses two. The output of all SA layers are concatenated in the feature dimension for both PCT and SPCT. The output of the last EdgeConv block is also added during concatenation with a skip connection. The detailed architectures used during training for PCT and SPCT are shown in Fig.~\ref{fig:PCT_arc}.

PCT and SPCT receive as inputs the particles found inside the jets. The input features vary between applications, depending on the available content for each public dataset. For all comparisons, up to 100 particles per jet are used. If more particles were found inside a jet, the event is truncated, otherwise zero-padded up to 100. The choice of 100 particles cover the majority of the jets stored in all datasets without truncation (more than 99\% of the cases for all datasets). To handle the zero-padded particles, the original PCT implementation has been modified such that these particles are masked during the k-nearest neighbors calculation and are replaced by large negative values to the matrix A defined in Eq. \ref{eq:A} such that the following softmax operation will result in zeros for zero-padded entries.

\subsection{HLS4ML LHC Jet dataset}
For this study, samples containing simulated jets originating from W bosons, Z bosons, top quarks, light quarks, and gluons  produced at $\sqrt{s}$ = 13 TeV proton-proton collisions are used. The samples are available at \cite{pierini_maurizio_2020_3602254}. This dataset is created and configured using a parametric description of a generic LHC detector, described in \cite{Coleman:2017fiq,Duarte:2018ite}. The jets are clustered with the anti-$k_{T}$ algorithm \cite{Cacciari:2008gp} with radius parameter R = 0.8, while also requiring that the jet's transverse momentum is around 1 TeV, which ensures that most of the decay products of the generated particles are found inside a single jet. 
  The training and testing sets contain 567k and 63k jets respectively. The performance comparison is reported using the official evaluation set, containing 240k jets.

For each jet, the momenta of all particle constituents are available. For each particle, a set of 16 kinematic features are used, matching the ones used in \cite{Moreno:2019bmu}, to facilitate the comparison. All input features available in the dataset are used without modification, with exception to the momenta and energy of the particles, which are taken after a logarithmic function is applied to limit the feature range. 

The area under the curve (AUC) for each class is calculated in the one-vs-rest strategy.  The results of the AUC are shown in Table \ref{tab:results_jedi} while the true positive rate (TPR) for a fixed false positive rate (FPR) at 10\% and 1\% are shown in Table \ref{tab:results_jedi_tpr}. PCT has the overall higher AUC compared to other algorithms, followed closely by SPCT and JEDI-net. For fixed FPR values, PCT also shows a good performance for all categories, achieving the best result for almost all categories and FPR values.

\begin{table}[htb]
    \centering
    \caption{Area under the curve and accuracy for each jet category reported on the HLS4ML LHC Jet dataset. Results for all algorithms are taken as the average of 10 trainings with random network initialization. If the uncertainty is not quoted then the variation is negligible compared to the expected value. Bold results represent the algorithm with highest performance. All results besides (S)PCT are taken from \cite{Moreno:2019bmu}.}
    \label{tab:results_jedi}
	\begin{tabular}{lcccccc}
    \hline
             Algorithm & Gluon & Light quark & W  boson & Z boson & Top quark\\
            \hline
            & &  AUC &  & & \\
            \hline
            DNN   & 0.9384 & 0.9026 & 0.9537 & 0.9459 & 0.9620 \\
            GRU  & 0.9040 & 0.8962 & 0.9192 & 0.9042 & 0.9350 \\
            CNN  & 0.8945  & 0.9007 & 0.9102 & 0.8994 & 0.9494\\
            JEDI-net & 0.9529 & 0.9301 & 0.9739& 0.9679 & 0.9683\\
            JEDI-net with $\sum O$   & 0.9528 & 0.9290 & 0.9695 & 0.9649 & 0.9677\\
            \hline
            SPCT & 0.9537 & 0.9326 & 0.9740 & 0.9779 & 0.9693 \\
            PCT & \textbf{0.9623} &  \textbf{0.9414} & \textbf{0.9789} & \textbf{0.9814}  & \textbf{0.9757} \\

	\end{tabular}
\end{table}

\begin{table}[htb]
    \centering
    \scriptsize
    \caption{True positive rate (TPR) for fixed values of the false positive rate (FPR). Results reported are the average of 10 independent trainings with random network initialization. Bold results represent the algorithm with largest TPR value. All results besides (S)PCT are taken from \cite{Moreno:2019bmu}.}
    \label{tab:results_jedi_tpr}
	\begin{tabular}{lcccccc}
    \hline
             Algorithm & Gluon & Light quark & W  boson & Z boson & Top quark\\
            \hline
            TPR for FPR=10\% &  & & \\
            \hline
            DNN   & $0.830 \pm 0.002$ & $0.715 \pm 0.002$ & $0.855 \pm 0.001$ & $0.833 \pm 0.002$ & $0.917 \pm 0.001$ \\
            GRU   & $0.740 \pm 0.014$ & $0.746 \pm 0.011$ & $0.812 \pm 0.035$ & $0.753 \pm 0.036$ & $0.867 \pm 0.006$ \\
            CNN   & $0.700 \pm 0.008$ & $0.740 \pm 0.003$ & $0.760 \pm 0.005$ & $0.721 \pm 0.006$ & $0.889 \pm 0.001$ \\
            JEDI-net & $0.878 \pm 0.001$ & $0.822 \pm 0.001$ & \textbf{0.938 $\pm$ 0.001} & $0.910 \pm 0.001$ & $0.930 \pm 0.001$ \\
            JEDI-net with $\sum O$  & $0.879 \pm 0.001$ & $0.818 \pm 0.001$ & $0.927 \pm 0.001$ & $0.903 \pm 0.001$ & $0.931 \pm 0.001$ \\
            \hline
            SPCT & $0.864 \pm 0.003$ & $0.807 \pm 0.002$ & $0.921 \pm 0.001$ & $0.940 \pm 0.002$ & $0.926 \pm 0.002$ \\
            PCT & \textbf{0.891 $\pm$ 0.001} & \textbf{0.833 $\pm$ 0.001} & $0.932 \pm 0.001$ & \textbf{0.946 $\pm$ 0.001} & \textbf{0.941 $\pm$ 0.001} \\
            \hline
            TPR for FPR=1\% &  & & \\
            \hline
            DNN   & $0.420 \pm 0.002$ & $0.178 \pm 0.002$ & $0.656 \pm 0.002$ & $0.715 \pm 0.001$ & $0.651 \pm 0.003$ \\
            GRU   & $0.273 \pm 0.018$ & $0.220 \pm 0.037$ & $0.249 \pm 0.057$ & $0.386 \pm 0.060$ & $0.426 \pm 0.020$ \\
            CNN   & $0.257 \pm 0.005$ & $0.254 \pm 0.007$ & $0.232 \pm 0.006$ & $0.291 \pm 0.005$ & $0.504 \pm 0.005$ \\
            JEDI-net & $0.485 \pm 0.001$ & \textbf{0.302 $\pm$ 0.001} & $0.704 \pm 0.001$ & $0.769 \pm 0.001$ & $0.633 \pm 0.001$ \\
            JEDI-net with $\sum O$  & $0.482 \pm 0.001$ & $0.301 \pm 0.001$ & $0.658 \pm 0.001$ & $0.729 \pm 0.001$ & $0.632 \pm 0.001$ \\
            \hline
            SPCT & $0.467 \pm 0.002$ & $0.283 \pm 0.002$ & $0.809 \pm 0.003$ & $0.753 \pm 0.002$ & $0.628 \pm 0.003$ \\
            PCT & \textbf{0.513 $\pm$ 0.002} & $0.298 \pm 0.002$ & \textbf{0.834 $\pm$ 0.001} & \textbf{0.781 $\pm$ 0.001} & \textbf{0.700 $\pm$ 0.003} \\ 
            
	\end{tabular}
\end{table}

\subsection{Top tagging dataset}
The top tagging dataset consists of jets containing the hadronic decay products of top quarks together with jets generated through QCD dijet events. The dataset is available at \cite{kasieczka_gregor_2019_2603256}. The events are generated with \textsc{Pythia8} \cite{Sjostrand:2014zea} with detector simulation done through \textsc{Delphes} \cite{deFavereau:2013fsa}. The jets are clustered with the anti-$k_{T}$ algorithm  with radius parameter R = 0.8. Only jets with transverse momentum $p_{T} \in [550,650] $ GeV and rapidity $|y|<2$ are kept. The official training, testing, and evaluation splitting are used, containing 1.2M/400k/400k events respectively.  For each particle, a set of 7 input features is used, based only on the momenta of each particle clustered inside the jet. The input features per particle are the same ones used in \cite{PhysRevD.101.056019} to facilitate the comparison between algorithms. Considering jets containing top quarks as the signal, the AUC and background rejection power, defined as the inverse of the background efficiency for a fixed signal efficiency, are listed in Tab.~\ref{tab:results_top}, with a reduced number of algorithms as reported in \cite{PhysRevD.101.056019}. A more complete, although slightly outdated list is available at \cite{Kasieczka:2019dbj}.

\begin{table}[htb]
    \centering
    \caption{Comparison between the performance reported for different classification algorithms on the top tagging dataset. The uncertainty quoted corresponds to the standard deviation of nine trainings with different random weight initialization. If the uncertainty is not quoted then the variation is negligible compared to the expected value. Bold results represent the algorithm with highest performance.}
    \label{tab:results_top}
	\begin{tabular}{lccccc}
          &  Acc &AUC & 1/$\epsilon_B$ ($\epsilon_S = 0.5$)  & 1/$\epsilon_B$ ($\epsilon_S = 0.3$) \\
            \hline
            ResNeXt-50 \cite{PhysRevD.101.056019} & 0.936 & 0.9837 & 302$\pm$5 & 1147$\pm$58 \\
            P-CNN \cite{PhysRevD.101.056019} & 0.930 & 0.9803 & 201$\pm$4 & 759$\pm$24 \\
            PFN \cite{Komiske_2019} & - & 0.9819 & 247$\pm$3 & 888$\pm$17 \\
            ParticleNet-Lite \cite{PhysRevD.101.056019} & 0.937 & 0.9844 &  325$\pm$5 & 1262$\pm$49 \\
            ParticleNet \cite{PhysRevD.101.056019} & \textbf{0.940} & \textbf{0.9858} & \textbf{397$\pm$7} & \textbf{1615$\pm$93}\\
            JEDI-net  \cite{Moreno:2019bmu}  & 0.9263 & 0.9786 & - & 590.4 \\
            JEDI-net with $\sum O$  \cite{Moreno:2019bmu}  & 0.9300 & 0.9807 & - & 774.6 \\
            \hline
            SPCT & 0.928 & 0.9799 & 201$\pm$9 & 725$\pm$54 \\
            PCT & \textbf{0.940} & 0.9855 & 392$\pm$7 & 1533$\pm$101 \\
	\end{tabular}
\end{table}

\subsection{Quark and gluon dataset}
The dataset used for the studies are available from \cite{Komiske_2019}, providing for each particle the same information present in the top tagging dataset (the momenta of the particles) with additional information that identifies the particle type as either electron, muon, charged hadron, neutral hadron, or photon. It consists of stable particles clustered into jets, excluding neutrinos, using the anti-$k_{T}$ algorithm with R = 0.4. The quark-initiated sample (treated as signal) is generated using a Z($\nu\nu$) + $(u,d,s)$ while the gluon-initiated data (treated as background) are generated using Z($\nu\nu$) +$g$ processes. Both samples are generated using \textsc{Pythia8} \cite{Sjostrand:2014zea} without detector effects. Jets are required to have transverse momentum $p_{T} \in [500,550]$ GeV and rapidity $|y|<1.7$ for the reconstruction. For the training, testing and evaluation, the  recommended splitting is used with 1.6M/200k/200k events respectively. Each particle contains the four momentum and the expected particles type (electron, muon, photon, or charged/neutral hadrons). For each particle, a set of 13 kinematic features is used. These features are chosen to match the ones used in \cite{PhysRevD.101.056019, Mikuni:2020wpr}. The AUC and background rejection power are listed in Tab.~\ref{tab:results_qg}.

\begin{table}[htb]
    \centering
    \caption{Comparison between the performance reported for different classification algorithms on the quark and gluon dataset. The uncertainty quoted corresponds to the standard deviation of nine trainings with different random weight initialization. If the uncertainty is not quoted then the variation is negligible compared to the expected value. Bold results represent the algorithm with best performance.}
    \label{tab:results_qg}
	\begin{tabular}{lccccc}
          &  Acc &AUC & 1/$\epsilon_B$ ($\epsilon_S = 0.5$)  & 1/$\epsilon_B$ ($\epsilon_S = 0.3$) \\
            \hline
            ResNeXt-50 \cite{PhysRevD.101.056019} & 0.821 & 0.9060 & 30.9 & 80.8 \\
            P-CNN \cite{PhysRevD.101.056019} & 0.827 & 0.9002 & 34.7 & 91.0 \\
            PFN \cite{Komiske_2019} & - & 0.9005 & 34.7$\pm$0.4 & - \\
            ParticleNet-Lite \cite{PhysRevD.101.056019} & 0.835 & 0.9079 &  37.1 & 94.5 \\
            ParticleNet \cite{PhysRevD.101.056019} & 0.840 & 0.9116 & 39.8$\pm$0.2 & 98.6$\pm$1.3\\
            ABCNet \cite{Mikuni:2020wpr}& 0.840 & 0.9126 & 42.6$\pm$0.4 & \textbf{118.4$\pm$1.5} \\
            \hline
            SPCT & 0.815 & 0.8910 & 31.6$\pm$0.3 & 93.0$\pm$1.2 \\
            PCT & \textbf{0.841} & \textbf{0.9140} & \textbf{43.2$\pm$0.7} & 118.0$\pm$2.2 \\
	\end{tabular}
\end{table}

\section{Ablation study \label{sec:ablation}}
To understand the benefits of the self-attention and the architecture choices for the development of (S)PCT applied to jet-tagging, we provide the comparison between the results attained by (S)PCT shown in the previous sections with different implementation choices. We start by comparing the benefit the SA layers bring to the classification by training the (S)PCT model without the SA layers, effectively taking the output of the feature extractor and passing it through the mean aggregation layer before the fully connected layers. While describing the PCT architecture, an offset-attention was used instead of the self-attention introduced in the original Transformer architecture. The comparison of both options has also been performed, where F$_{sa}$, introduced in Eq. \ref{eq:fsa}, is used as the output of the SA layer. These comparisons are summarized in Tables \ref{tab:ablation_jedi_auc} and \ref{tab:ablation_jedi_eff} with the HLS4ML LHC Jet dataset,  Table \ref{tab:ablation_top} with the top tagging dataset, and \ref{tab:ablation_qg} with the quark and gluon dataset. The values reported for the different scenarios  are the result of a single training (hence no uncertainties provided), but is has been verified that  the results shown are stable with additional trainings.

\begin{table}[htb]
    \centering
    \scriptsize
    \caption{Area under the curve for each jet category of the HLS4ML LHC Jet dataset for different implementations of the (S)PCT architecture. Values in parenthesis represent the percentage difference with the standard implementation.}
    \label{tab:ablation_jedi_auc}
	\begin{tabular}{lcccccc}
    \hline
             Algorithm & Gluon & Light quark & W  boson & Z boson & Top quark\\
            \hline
            & & &  AUC  & & \\
            \hline
            SPCT & 0.9537 & 0.9326 & 0.9740 & 0.9779 & 0.9693 \\
            SPCT w/o attention & 0.9399 (-1.4\%) & 0.918 (-1.6\%) & 0.9501 (-2.5\%) & 0.9641 (-1.4\%) & 0.956 (-1.4\%)) \\
            SPCT w/o offset-attention &0.9543 (+0.1\%) & 0.932 (-0.1\%) & 0.9737 (-0.0\%) & 0.9772 (-0.1\%) & 0.9691 (-0.0\%) \\
            \hline
            PCT  & 0.9623 &  0.9414 & 0.9789 & 0.9814  & 0.9757 \\
            PCT w/o attention &0.9578 (-0.5\%) & 0.9362 (-0.6\%) & 0.9736 (-0.5\%) & 0.9774 (-0.4\%) & 0.9707 (-0.5\%) \\
            PCT w/o offset-attention & 0.9615 (-0.1\%) & 0.9404 (-0.1\%) & 0.9785 (-0.0\%) & 0.9812 (-0.0\%) & 0.9749 (-0.1\%) \\

	\end{tabular}
\end{table}

\begin{table}[htb]
    \centering

    \caption{True positive rate (TPR) for fixed values of the false positive rate (FPR) for different implementations of the (S)PCT architecture evaluated in the HLS4ML LHC Jet dataset. Values in parenthesis represent the percentage difference with the standard implementation.}
    \scriptsize
    \label{tab:ablation_jedi_eff}
	\begin{tabular}{lcccccc}
    \hline
             Algorithm & Gluon & Light quark & W  boson & Z boson & Top quark\\
            \hline
            TPR for FPR=10\% &  & & \\
            \hline
            SPCT & $0.864$ & $0.807$ & $0.921$ & $0.940$ & $0.926$ \\
            SPCT w/o attention &0.826 (-4.4\%) & 0.775 (-4.0\%) & 0.82 (-11.0\%) & 0.911 (-3.1\%) & 0.902 (-2.6\%) \\
            SPCT w/o offset-attention &0.866 (+0.2\%) & 0.805 (-0.2\%) & 0.92 (-0.1\%) & 0.938 (-0.2\%) & 0.924 (-0.2\%)  \\
            \hline
            PCT & 0.891  & 0.833  & $0.932 $ & 0.946  & 0.941  \\
            PCT w/o attention & 0.878 (-1.5\%) & 0.819 (-1.7\%) & 0.922 (-1.1\%) & 0.937 (-1.0\%) & 0.929 (-1.3\%)  \\
            PCT w/o offset-attention & 0.867 (-2.7\%) & 0.829 (-0.5\%) & 0.932 (-0.0\%) & 0.946 (-0.0\%) & 0.939 (-0.2\%)  \\
            \hline
            TPR for FPR=1\% &  & & \\
            \hline
            SPCT & $0.467$ & $0.283$ & 0.809 & $0.753 $ & $0.628$ \\
            SPCT w/o attention & 0.422 (-9.6\%) & 0.263 (-7.1\%) & 0.697 (-13.8\%) & 0.594 (-21.1\%) & 0.51 (-18.8\%)  \\
            SPCT w/o offset-attention &0.469 (+0.4\%) & 0.282 (-0.4\%) & 0.806 (-0.4\%) & 0.747 (-0.8\%) & 0.63 (+0.3\%)  \\
            \hline
            PCT & 0.513  & $0.298 $ & 0.834  & 0.781  & 0.700  \\ 
            PCT w/o attention &0.493 (-3.9\%) & 0.289 (-3.0\%) & 0.797 (-4.4\%) & 0.745 (-4.6\%) & 0.648 (-7.4\%)  \\
            PCT w/o offset-attention & 0.507 (-1.2\%) & 0.297 (-0.3\%) & 0.833 (+0.1\%) & 0.782 (--0.1\%) & 0.683 (-2.4\%)  \\
            
	\end{tabular}
\end{table}

\begin{table}[htb]
    \centering
    \footnotesize
    \caption{Performance comparison for different (S)PCT implementations on the top tagging dataset. Values in parenthesis represent the percentage difference with the standard implementation.}
    \label{tab:ablation_top}
	\begin{tabular}{lccccc}
          &  Acc &AUC & 1/$\epsilon_B$ ($\epsilon_S = 0.5$)  & 1/$\epsilon_B$ ($\epsilon_S = 0.3$) \\
            \hline
            SPCT & 0.928 & 0.9799 & 201 & 725 \\
            SPCT w/o attention & 0.91 (-1.9\%) & 0.964 (-1.6\%) & 52 (-74.1\%) & 130 (-82.1\%)    \\
            SPCT w/o offset-attention & 0.926 (-0.2\%) & 0.9789 (-0.1\%) & 183 (-9.0\%) & 645 (-11.0\%)  \\
            \hline
            PCT & 0.940 & 0.9855 & 392 & 1533 \\
            PCT w/o attention &0.938 (-0.2\%) & 0.9846 (-0.1\%) & 330 (-15.8\%) & 1177 (-23.2\%)    \\
            PCT w/o offset-attention & 0.939 (-0.1\%) & 0.9849 (-0.1\%) & 350 (-10.7\%) & 1343 (-12.4\%)  \\
	\end{tabular}
\end{table}

\begin{table}[htb]
    \centering
    \footnotesize
    \caption{Performance comparison for different (S)PCT implementations on the quark gluon dataset. Values in parenthesis represent the percentage difference with the standard implementation.}
    \label{tab:ablation_qg}
	\begin{tabular}{lccccc}
          &  Acc &AUC & 1/$\epsilon_B$ ($\epsilon_S = 0.5$)  & 1/$\epsilon_B$ ($\epsilon_S = 0.3$) \\
            \hline
            SPCT & 0.815 & 0.8910 & 31.6 & 93.0 \\
            SPCT w/o attention & 0.803 (-1.5\%) & 0.8786 (-1.4\%) & 27.1 (-14.2\%) & 79.8 (-14.2\%)   \\
            SPCT w/o offset-attention & 0.801 (-1.7\%) & 0.8765 (-1.6\%) & 26.9 (-14.9\%) & 78.5 (-15.6\%)  \\
            \hline
            PCT & 0.841 & 0.9140 & 43.2& 118.0 \\
            PCT w/o attention & 0.826 (-1.8\%) & 0.9012 (-1.4\%) & 34.0 (-21.3\%) & 97.4 (-17.5\%)  \\
            PCT w/o offset-attention & 0.823 (-2.1\%) & 0.898 (-1.8\%) & 31.6 (-26.9\%) & 86.9 (-26.4\%)   \\
	\end{tabular}
\end{table}

The implementation of PCT without the attention modules results in an architecture that is similar to the ParticleNet-Lite implementation from \cite{PhysRevD.101.056019}, also resulting in similar performance for both top tagging and quark gluon datasets. The differences for SPCT with and without the attention modules are much bigger compared to PCT. Without the attention layers, SPCT does not have access to the information shared between particles, resulting in worse performance. Since for PCT the neighboring information is already used in the feature extractor, this effect is mitigated. Compared to the original self-attention implementation, the offset-attention shows better performance, which is more noticeable in the quark gluon dataset. The quark gluon dataset uses the same input features per particle as the top tagging dataset, but with an additional particle identifier (PID) information, encoded as binary entries to categorize the particles into five categories: muons, electrons, photons, charged hadrons, and neutral hadrons. Given that results without attention for the quark gluon dataset also seem to yield better performance compared to the standard attention, this might indicate that binary input features are not correctly handled by the standard attention implementation. In the offset-attention, the module calculates an offset of the inputs per particle, relying less on the absolute value of the PID flag.

\section{Computational complexity}
Besides the algorithm performance, the computational cost is also an important figure of merit. To compare the amount of computational resources required to evaluate each model, the number of trainable weights and the number of floating point operations (FLOPs) are computed. The comparison of these quantities for different algorithms are shown in Tab.~\ref{tab:resources}.

\begin{table}[htb]
    \centering
    \caption{Number of trainable weights and floating point operations (FLOPs) for each model under consideration. Missing entries are left for implementations that do not report an official value.}
    \label{tab:resources}
	\begin{tabular}{lccc}

             Algorithm & Weights & FLOPs\\
            \hline
            ResNeXt-50 \cite{PhysRevD.101.056019} & 1.46M & -\\
            P-CNN  \cite{PhysRevD.101.056019} & 348k & - \\
            PFN   \cite{Komiske_2019} & 82k & - \\
            ParticleNet-Lite  \cite{PhysRevD.101.056019} & 26k & -\\
            ParticleNet  \cite{PhysRevD.101.056019} & 366k & -\\
            ABCNet \cite{Mikuni:2020wpr} & 230k & -\\
            DNN  \cite{Moreno:2019bmu} & 14.7k & 27k \\
            GRU  \cite{Moreno:2019bmu}  & 15.6k & 46k \\
            CNN  \cite{Moreno:2019bmu}  & 205.5k & 400k \\
            JEDI-net  \cite{Moreno:2019bmu}  & 33.6k & 116M\\
            JEDI-net with $\sum O$  \cite{Moreno:2019bmu}  & 8.8k & 458M\\
            \hline
            SPCT & 7.0k & 2.4M \\
            PCT & 193.3k &  266M \\
	\end{tabular}
\end{table}

While PCT shows a better overall AUC compared to SPCT, the improvement in performance from the usage of EdgeConv blocks comes with a cost in computational complexity. SPCT, on the other hand, provides a good balance between performance and computational cost, resulting in more than 100 times less FLOPs and almost 30 times less trainable weights compared to PCT. Although JEDI-net shows similar performance to SPCT, the implementation takes as inputs 150 particles with the model based on $\bar{O}$ sums appearing with a large FLOP count of 458M. This number however can be decreased with a sparse implementation, since many of these operations are $\times 0$ and $\times 1$ products.

\section{Visualization}
\begin{figure}[htb]
    \centering
    \includegraphics[width=0.18\textwidth]{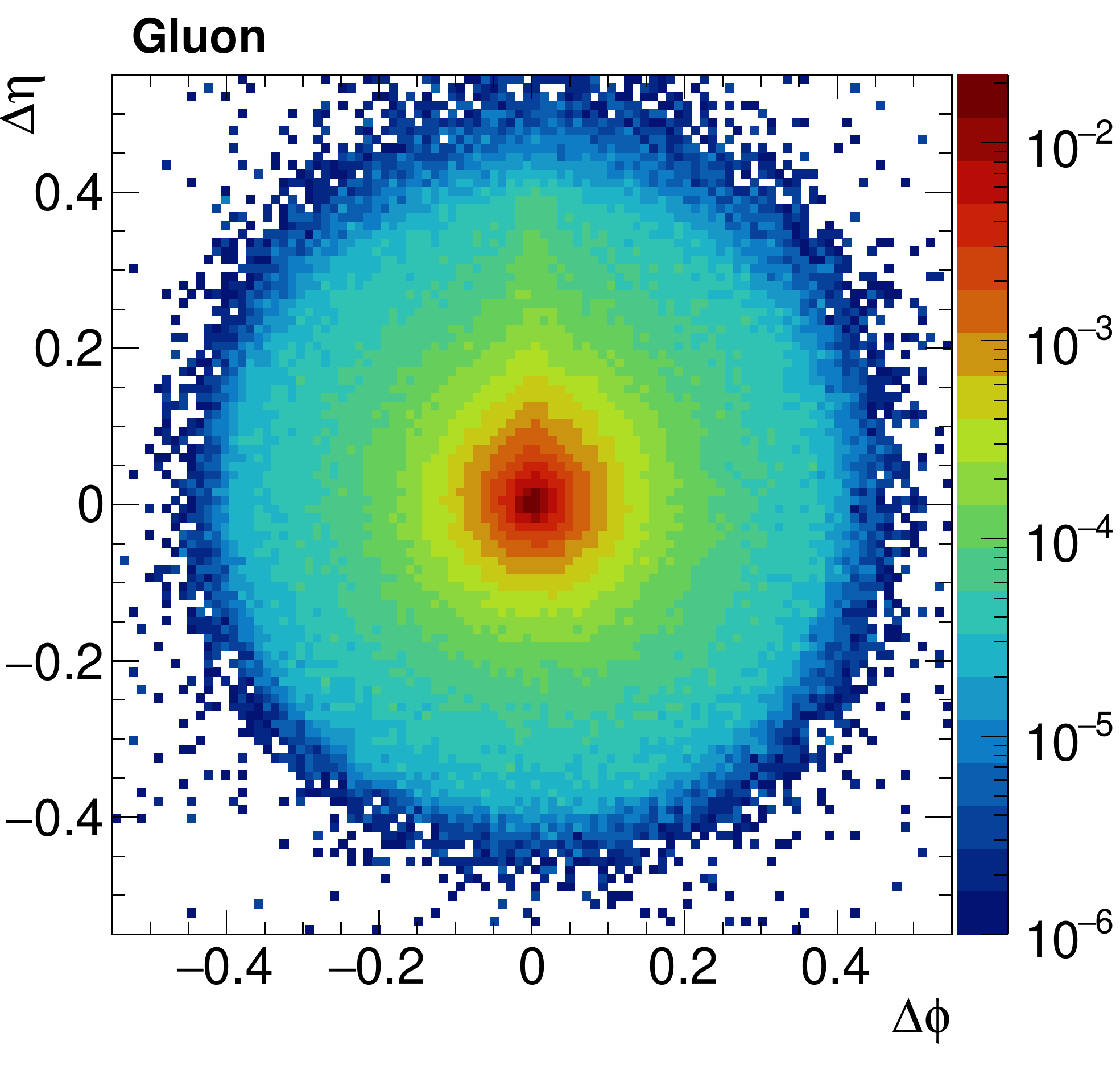}
    \includegraphics[width=0.18\textwidth]{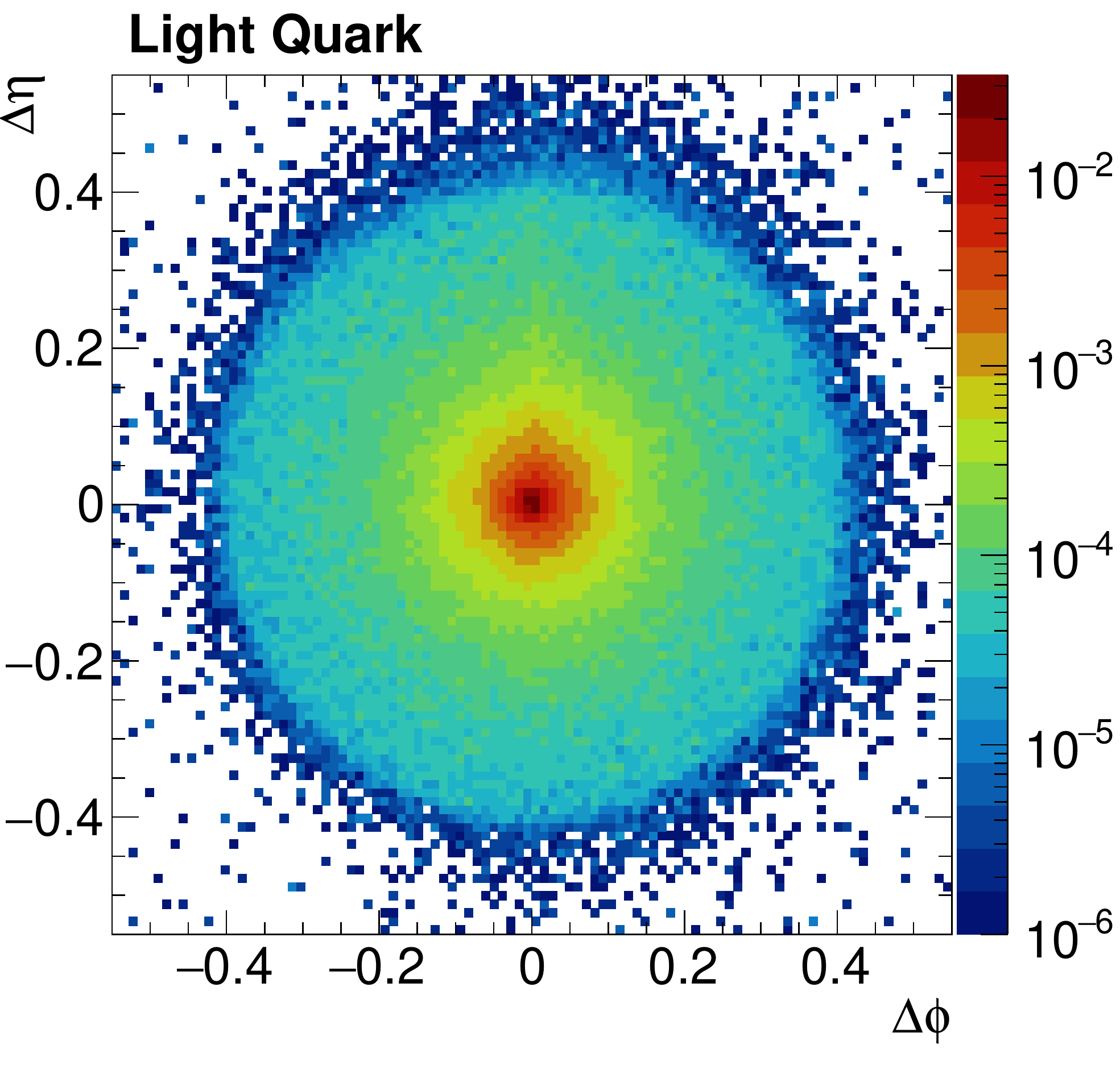}
    \includegraphics[width=0.18\textwidth]{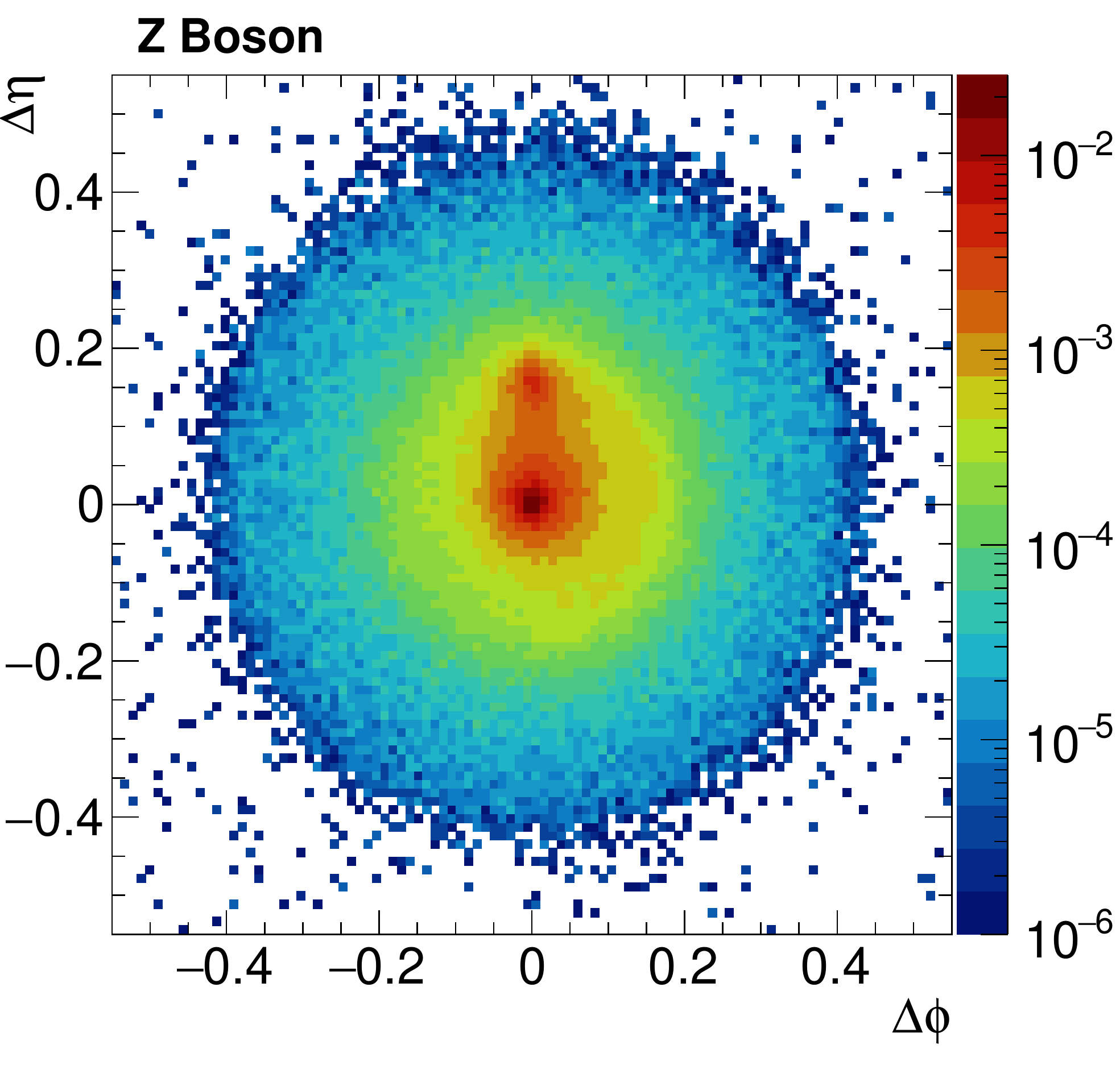}
    \includegraphics[width=0.18\textwidth]{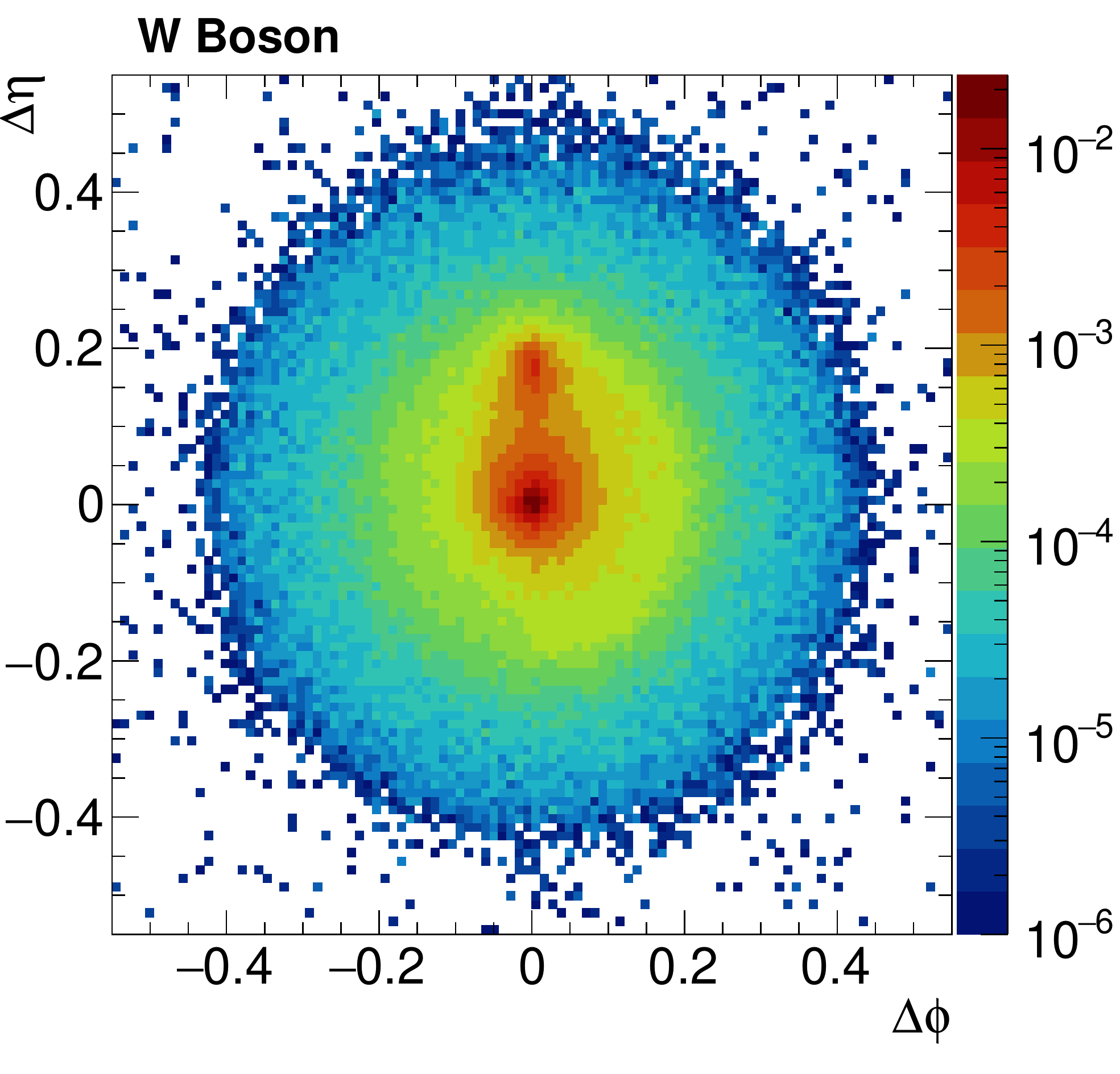}
    \includegraphics[width=0.18\textwidth]{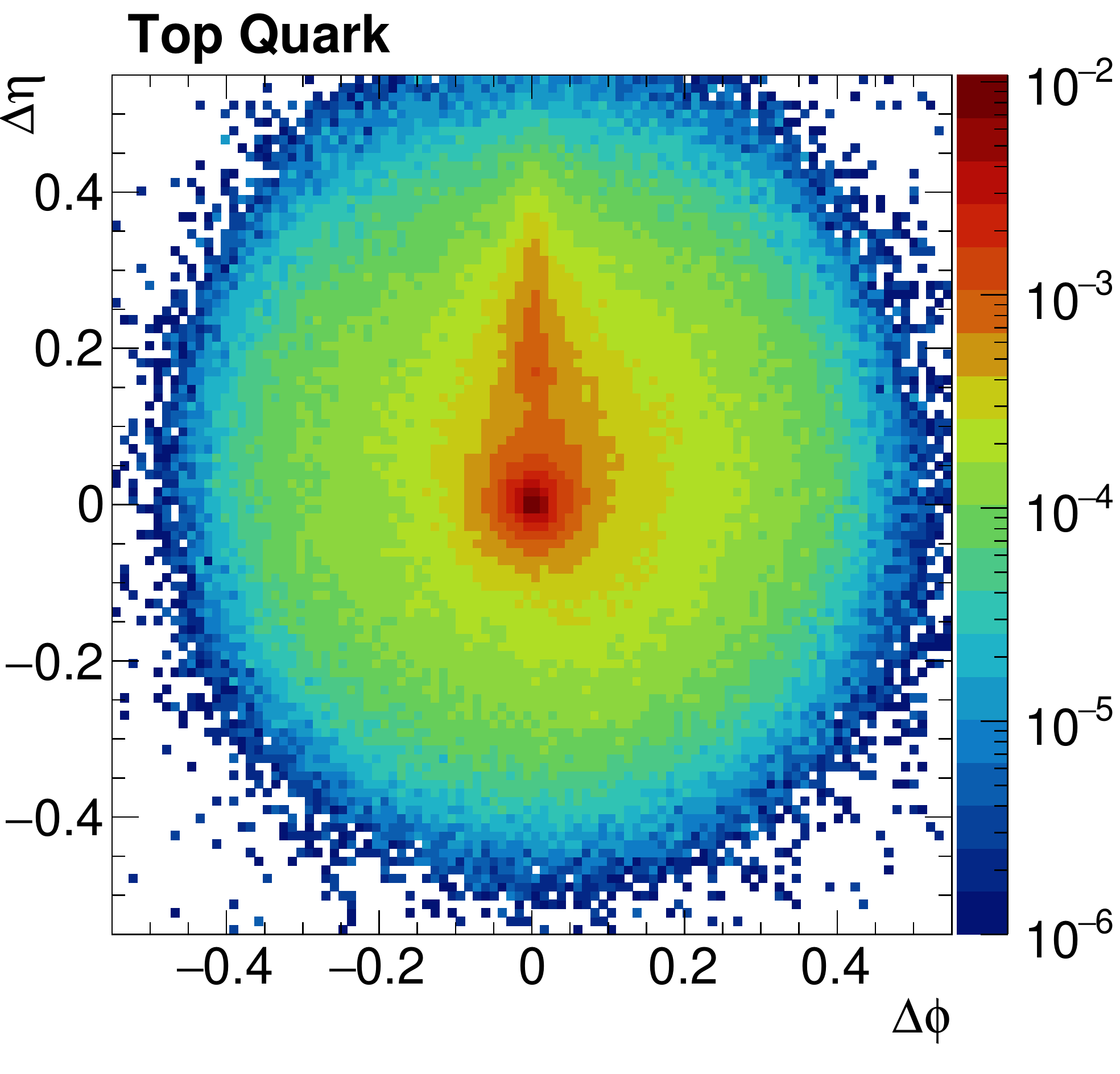}
    
    \includegraphics[width=0.18\textwidth]{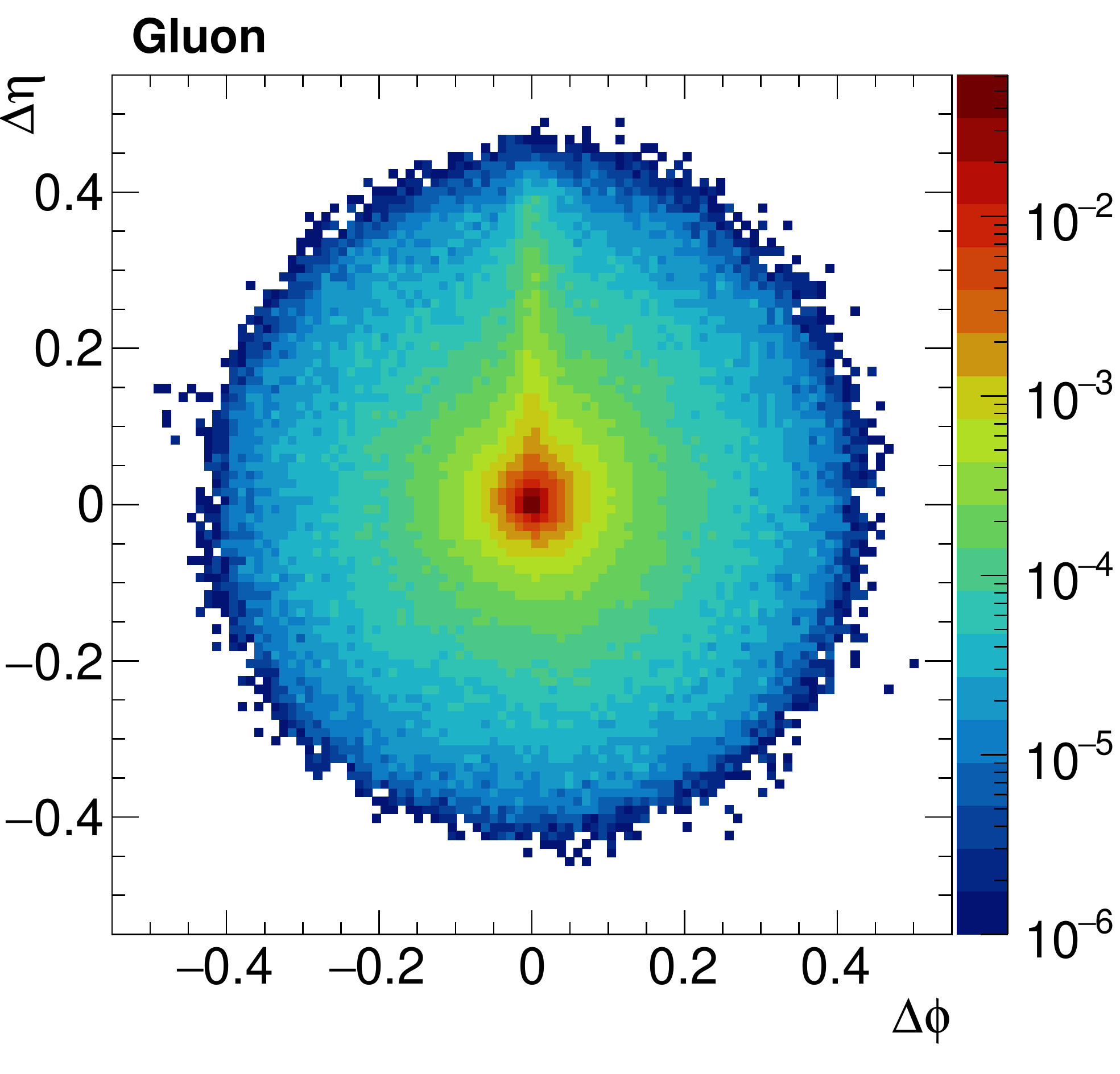}
    \includegraphics[width=0.18\textwidth]{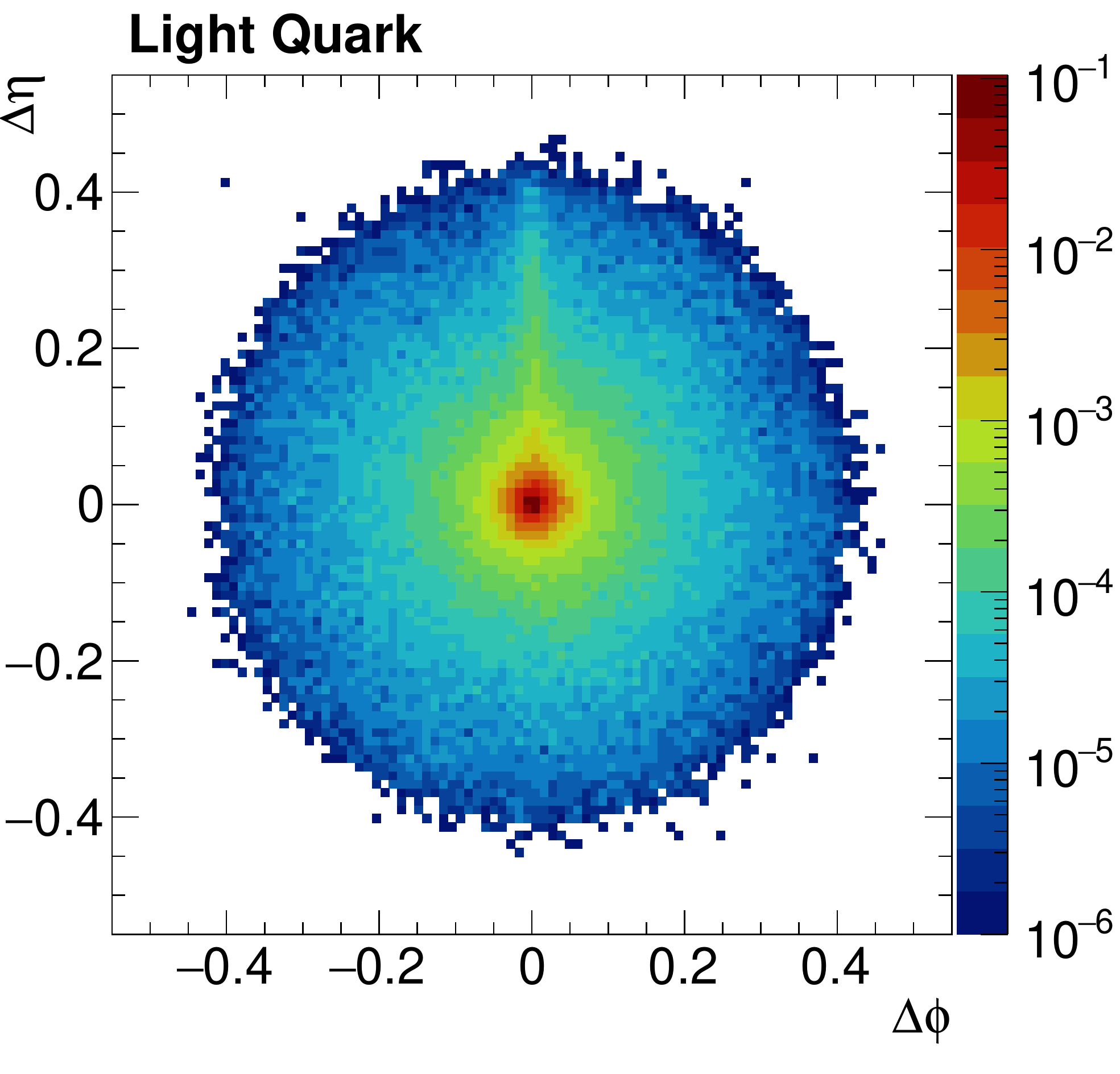}
    \includegraphics[width=0.18\textwidth]{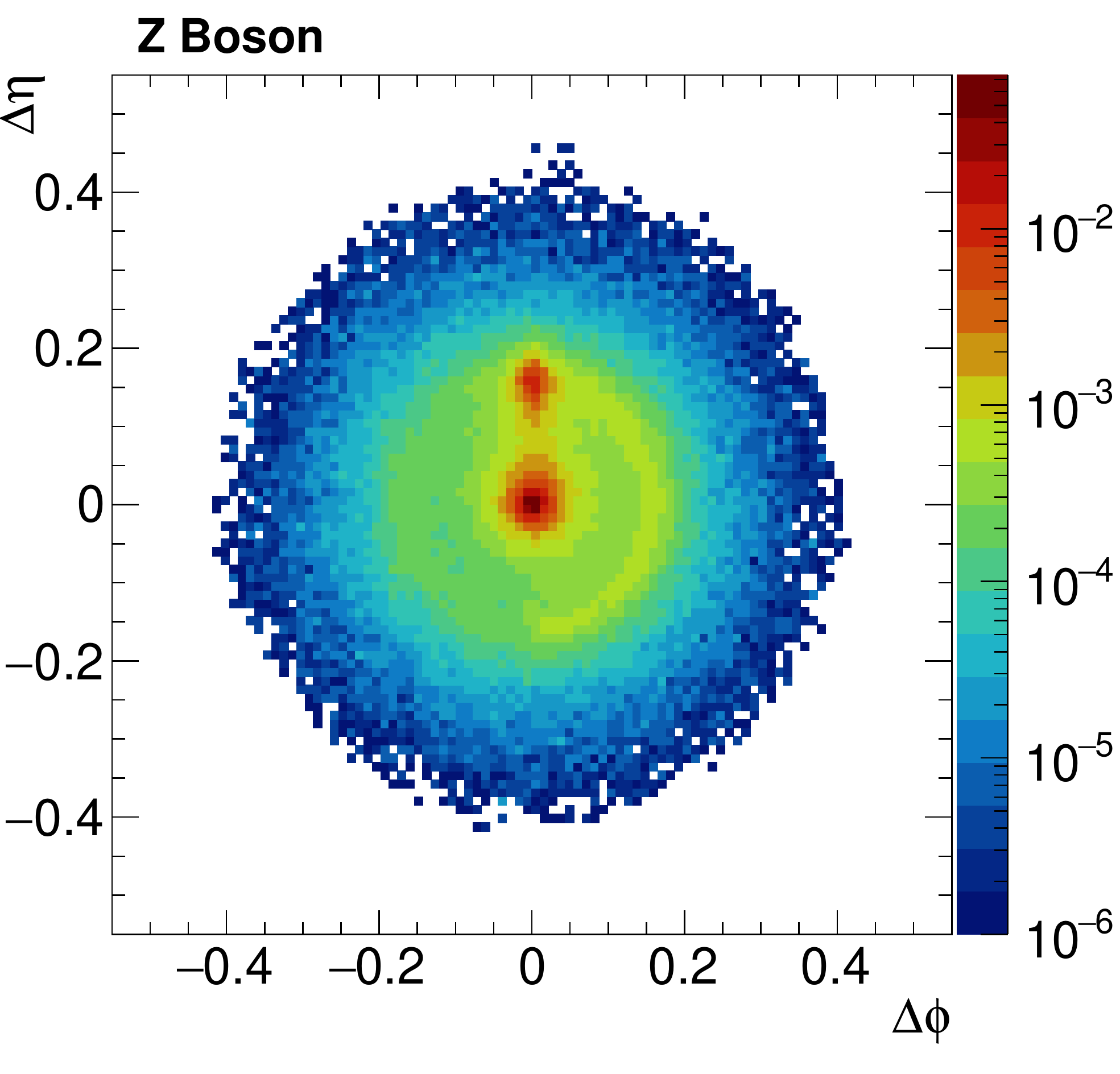}
    \includegraphics[width=0.18\textwidth]{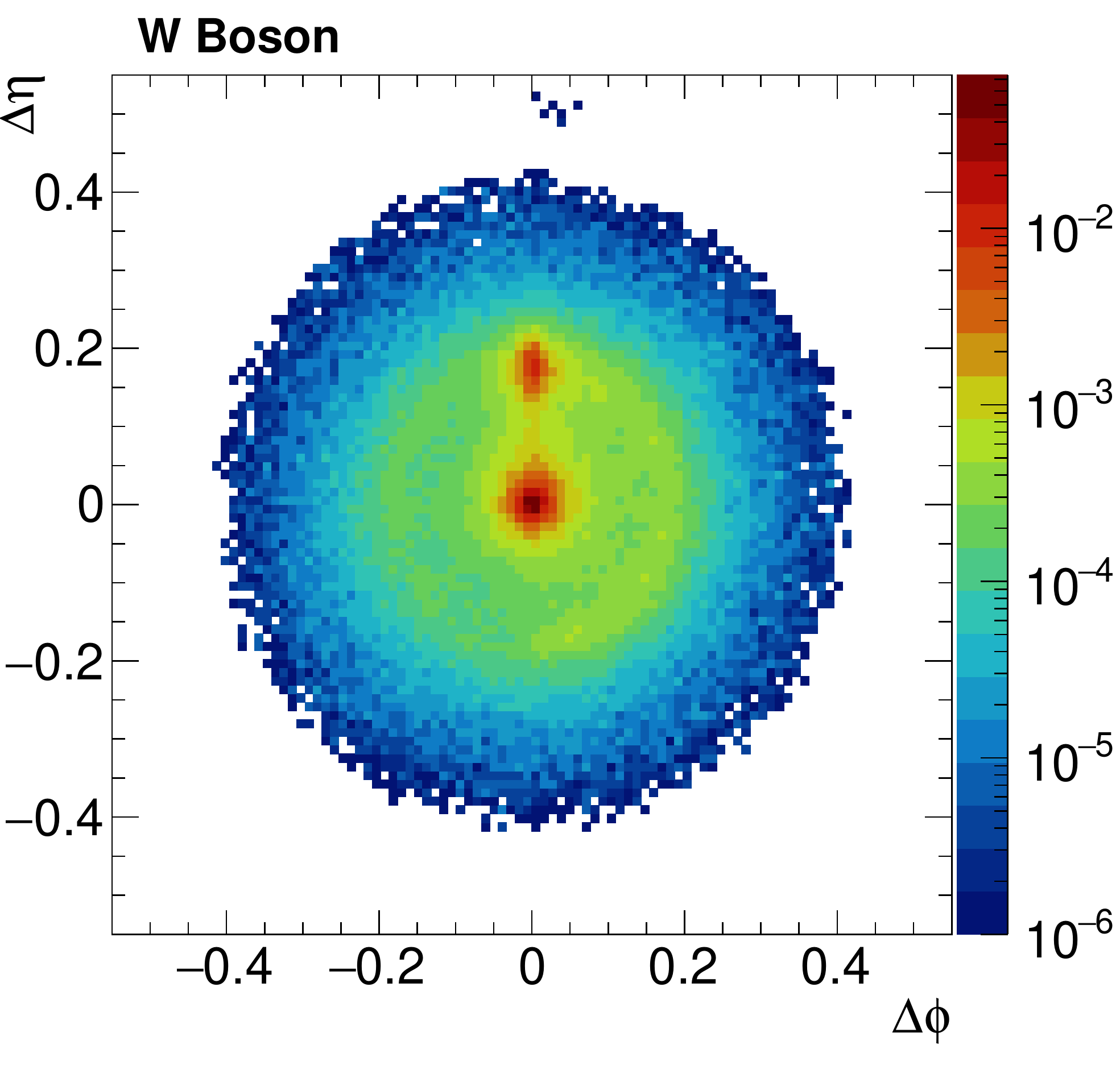}
    \includegraphics[width=0.18\textwidth]{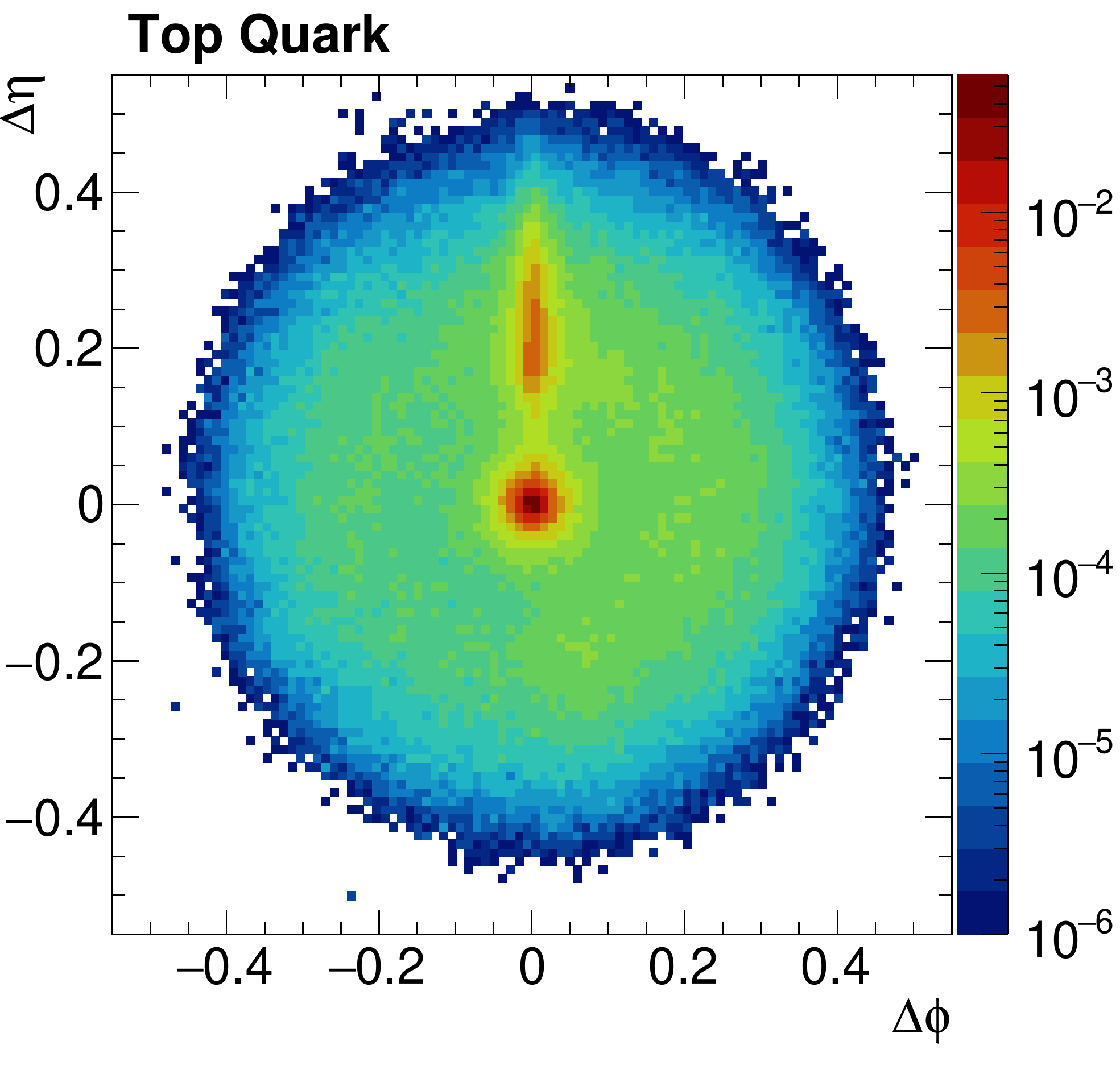}
    
    \includegraphics[width=0.18\textwidth]{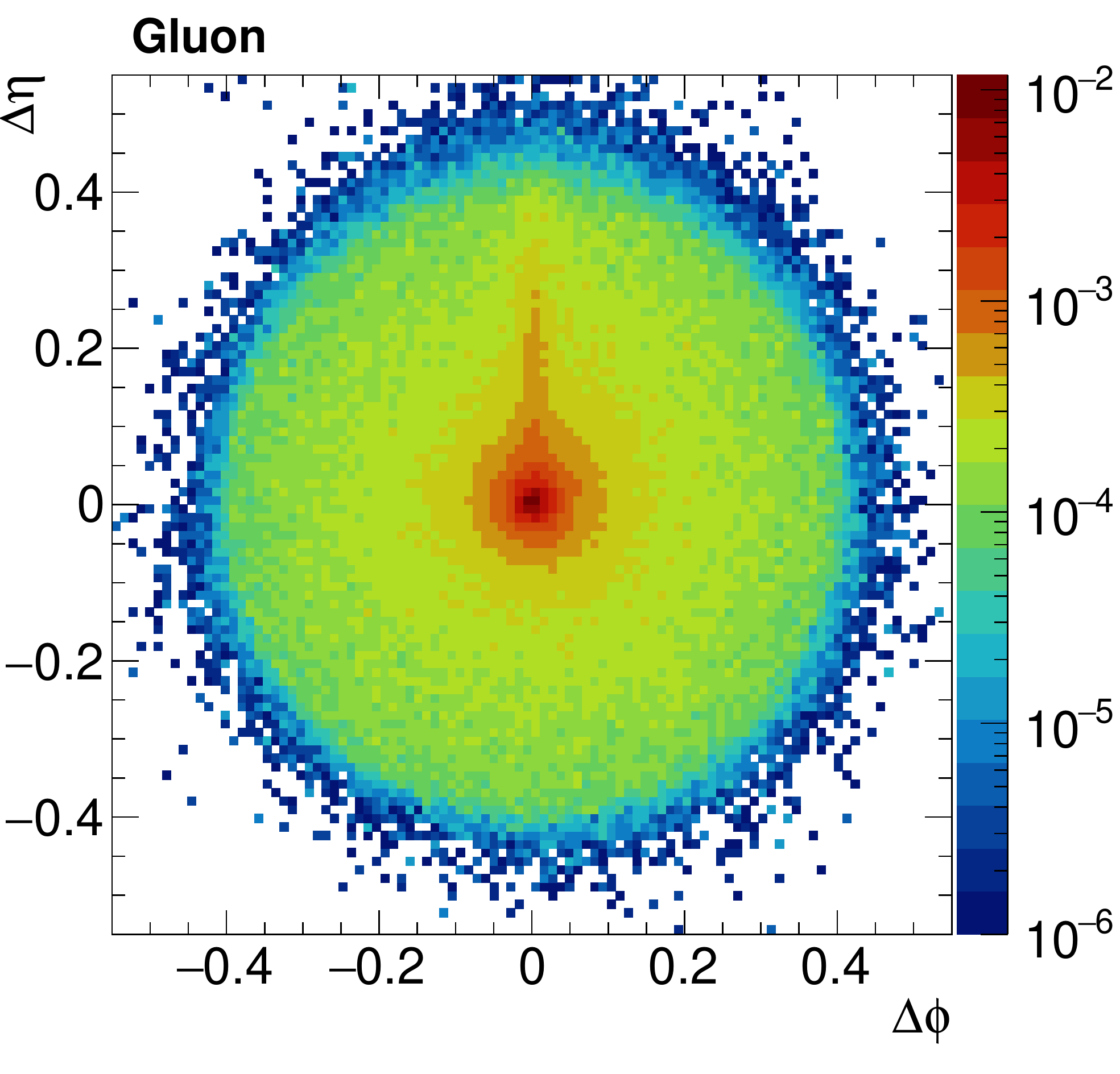}
    \includegraphics[width=0.18\textwidth]{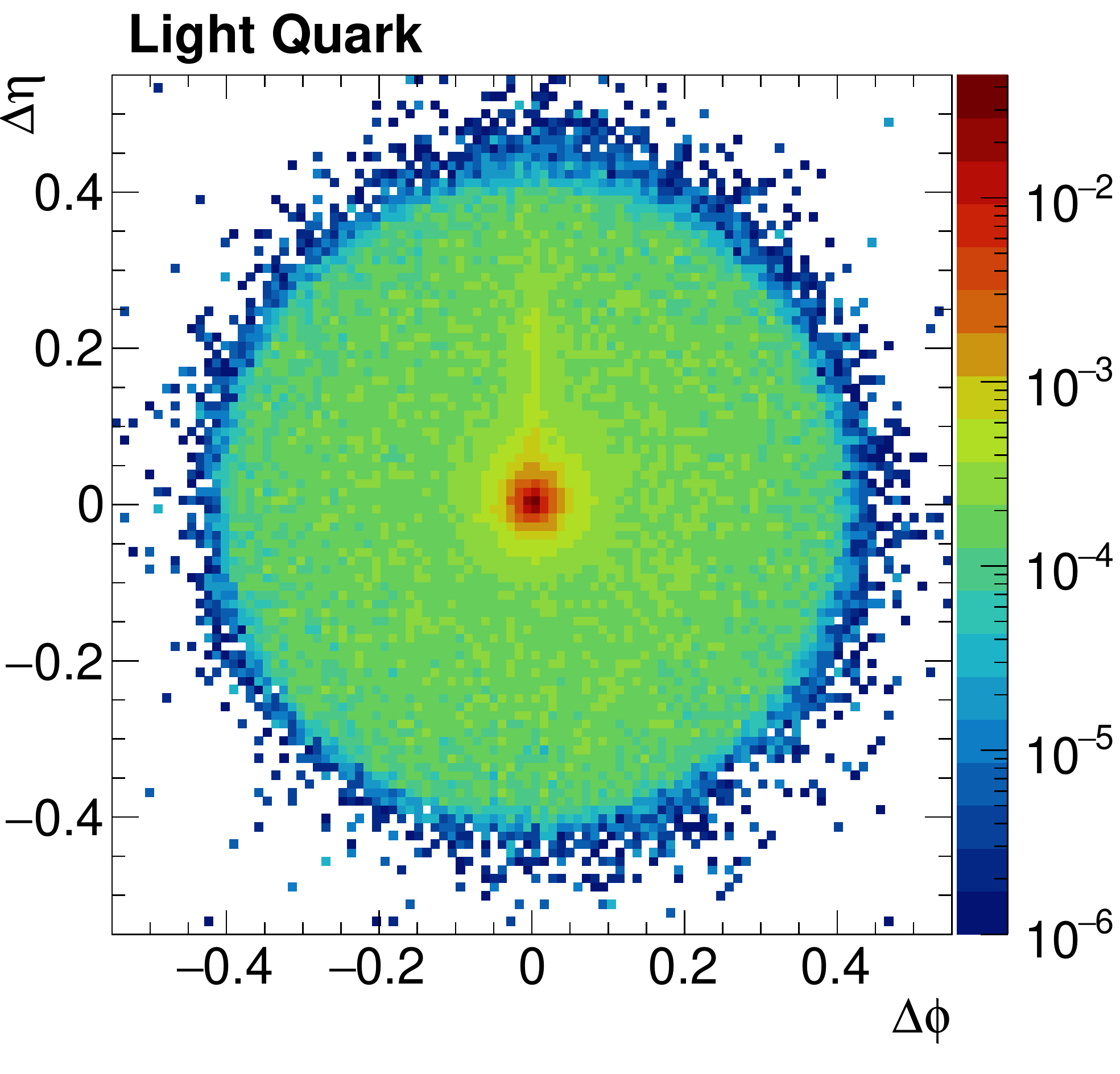}
    \includegraphics[width=0.18\textwidth]{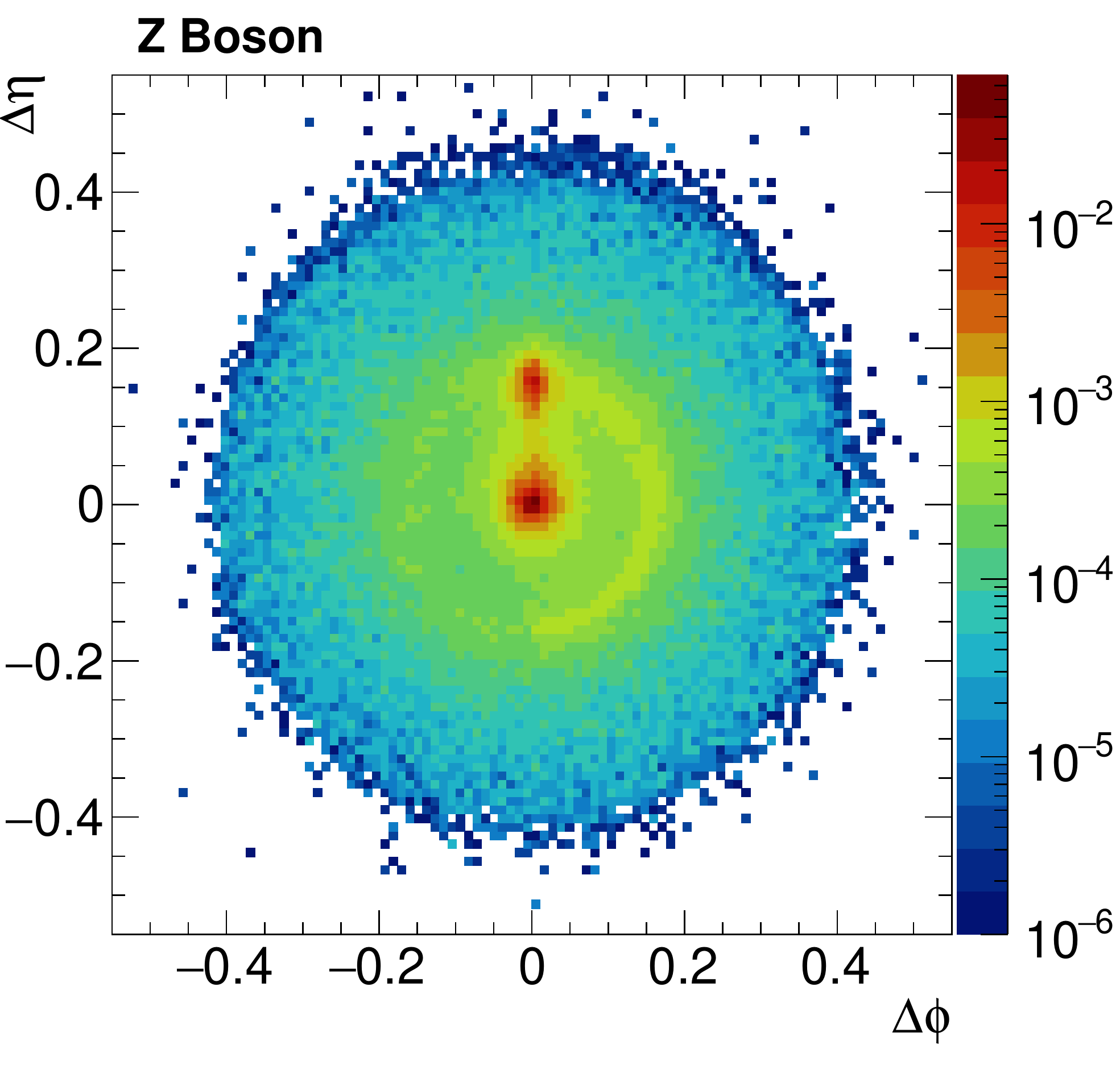}
    \includegraphics[width=0.18\textwidth]{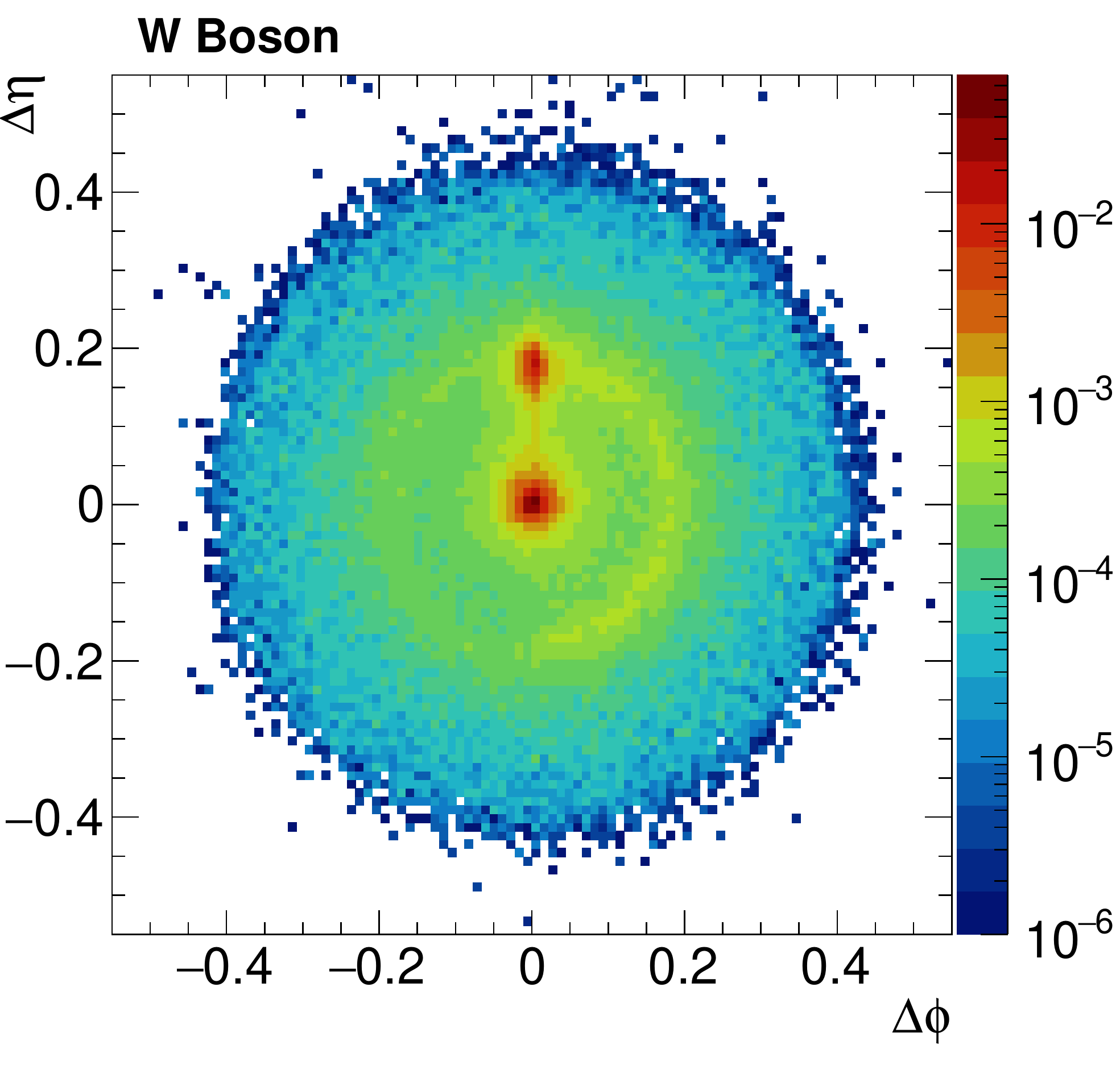}
    \includegraphics[width=0.18\textwidth]{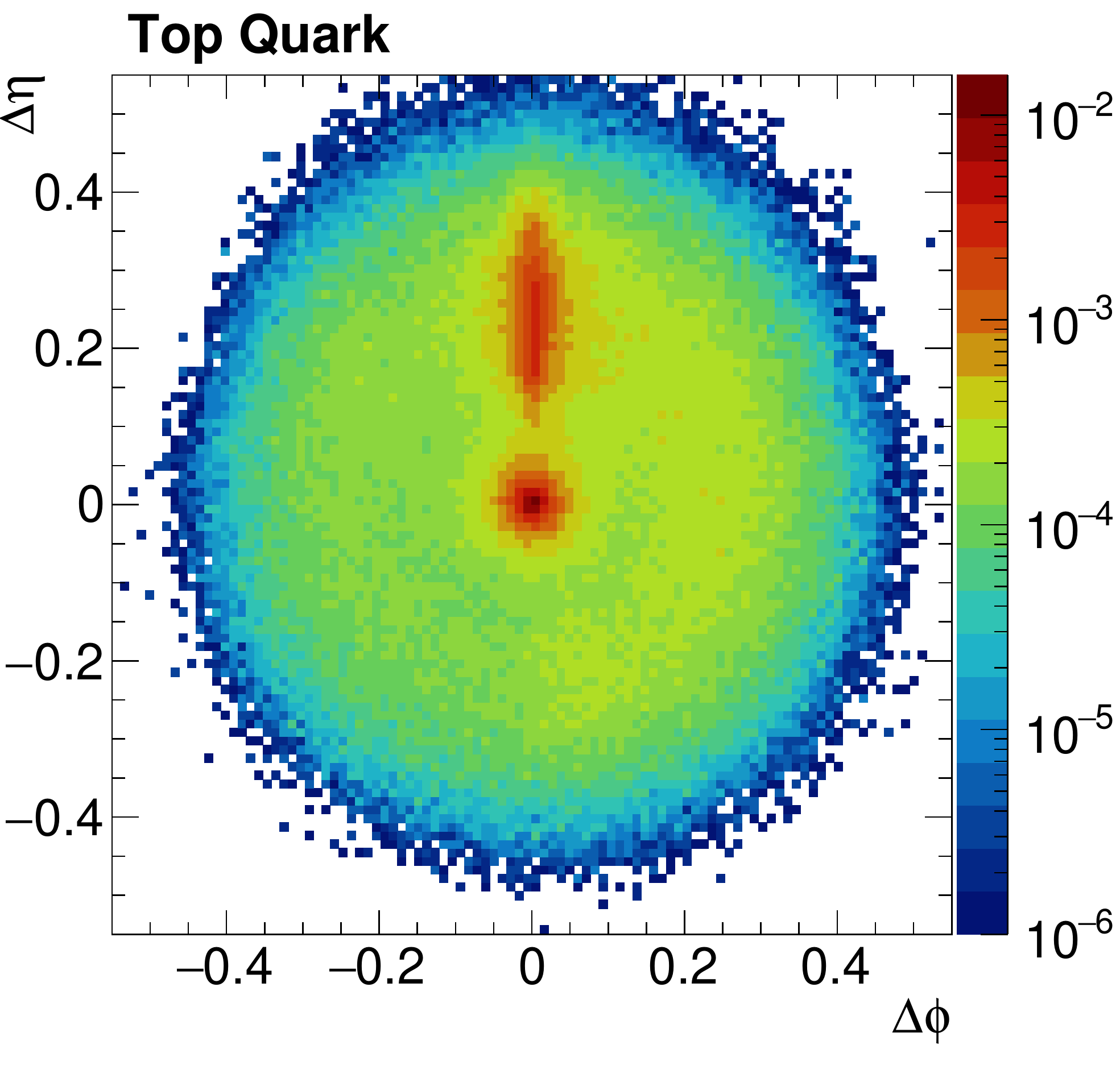}

    \caption{Average jet image for each jet category (columns) and for each self-attention layer (rows). The pixel intensities represent the overall particle importance compared to the most energetic particle in the jet.}
    \label{fig:att}
\end{figure}

The SA module defines the relative importance between all points in the set through the attention weights. We can use this information to identify the regions inside a jet that have high importance for a chosen particle. To visualize the particle importance, the HLS4ML LHC jet dataset is used to create a pixelated image of a jet in the transverse plane. The average jet image of 100k examples in the evaluation set is used. For each image, a simple preprocessing strategy is applied to align the different images. First, the whole jet is translated such that the particle with the highest transverse momentum in the jet is centered at (0,0). This particle is also used as the reference particle from where attention weights are shown. Next, the full jet image is rotated, making the second most energetic particle aligned with the positive y-coordinate. Lastly, the image is flipped in the x-coordinate in case the third most energetic particle is located on the negative x-axis, otherwise the image is left as is. These transformations are also used in other jet image studies such as \cite{Komiske:2016rsd,Mikuni:2020wpr}. The pixel intensity for each jet image is taken from the attention weights after the softmax operation is applied, expressing the particle importance with respect to the most energetic particle in the event. A comparison of the extracted images for each SA layer and for each jet category is shown in Fig.~\ref{fig:att} .

The different SA layers are able to extract different information for each jet. In particular, the jet substructure is exploited, resulting in an increased relevance to harder subjets in the case of Z boson, W boson, and top quark initiated jets. On the other hand, light quark and gluon initiated jets have a more homogeneous radiation pattern, resulting also in a more homogeneous picture. 

\section{Summary of experimental results}
In this section, the summary of the comparisons made using different datasets is given. In the HLS4ML dataset, both PCT and SPCT show state-of-the-art performance, resulting in superior AUC compared to other approaches. Although the AUC provides a general idea of the performance, the TPR at fixed FPR thresholds represent a more realistic application of the algorithm in high energy physics, where one is interested in maximizing TPR while having control of the background level. in these comparisons again PCT and SPCT show a superior performance for the majority of the jet categories. In the top tagging and quark gluon datasets, a similar picture is shown, with PCT and SPCT again performing well compared to other algorithms. In the top tagging dataset, the results achieved with PCT and ParticleNet are similar and compatible within training uncertainties. This observation suggests that the information encoded through EdgeConv layers might already provide enough model abstraction to perform the classification. On the other hand, in the quark gluon dataset, the performance of PCT compared to ParticleNet shows a 20\% improvement in background rejection power at 30\% signal efficiency. This performance is similar to the one reported by the ABCNet architecture that implements attention through graph attention pooling layers \cite{DBLP:conf/iclr/VelickovicCCRLB18}, evidencing some of the benefits of the usage of attention mechanisms.

\section{Conclusion}
In this work, a new method based on the Transformer architecture was applied to a high energy physics application. The point cloud transformer (PCT) modifies the usual Transformer architecture to be applied to a set of unordered points present in a point cloud. This method has the advantage of extracting semantic affinities between the points through the development of a self-attention mechanism. We evaluate the performance of this architecture applied to several jet-tagging datasets by testing two different implementations, one that exploits the neighborhood information through EdgeConv operations and a simpler form that connects all points through  convolutional layers called simple PCT (SPCT). Both approaches have shown state-of-the-art performance compared to other publicly available results. While the classification performance of SPCT is slightly lower compared to the standard PCT, the number of floating point operations required to evaluate the model decreases by almost a factor 20. This reduced computational complexity can be exploited in environments with limited computing resources or applications that require fast inference responses.

A different advantage of (S)PCT is the visualization of the self-attention coefficients to understand which points have a greater importance through the classification task. Traditional methods often define physics-motivated observables to distinguish the different types of jets. PCT, on the other hand, exploits subjet information by learning affinities on a particle-by-particle basis, resulting in images with distinct features for jets of different decay modes. 

\section{Acknowledgements}
The authors would like to thank Jean-Roch Vlimant for helpful comments during the development of this work.
This research was supported in part by the Swiss National Science Foundation (SNF) under contract No. 200020-182037 and Forschungskredit of the Universityof Zurich, grant no. FK-20-097. 

\section{References} 
\bibliographystyle{JHEP}
\bibliography{ref}

\end{document}